\documentclass{webofc}
\usepackage{graphicx}

\usepackage[varg]{txfonts}   
\usepackage{hyperref}
\usepackage{url}
\usepackage{lineno}
\usepackage[T1]{fontenc}
\usepackage[utf8]{inputenc} 
\usepackage{textcomp}
\hypersetup{colorlinks=true,citecolor=blue,urlcolor=blue,linkcolor=blue}
\begin{document}
%
\title{Probing the strong interaction between charm hadrons and charged particles with femtoscopy measurements with \mbox{ALICE}
}
%
%

\author{\firstname{Biao} \lastname{Zhang}\inst{1}\fnsep\thanks{\email{biao.zhang@cern.ch}} 
, on behalf of the ALICE Collaboration}

\institute{Physikalisches Institut, Universität Heidelberg}

\abstract{Studies of strong interactions between hadrons provide a valuable opportunity to test Quantum Chromodynamics at nucleon-scale distances. The femtoscopy technique has proven to be an effective tool for studying interactions between unstable hadrons by measuring the correlation function of hadron pairs in momentum space. While several measurements of the strong interactions between light and strange hadrons have been conducted using this technique, studies of charm hadrons have been limited. These studies can shed light on the formation of exotic charm states or, in the case of baryons, nuclei with charm content.

In these proceedings, measurements of the strong interaction between charm hadrons and light-flavor hadrons using the femtoscopy technique are presented. The final results on the correlation functions and residual strong interactions between $\rm D$ mesons and light hadrons measured in pp collisions at $\sqrt{s} = 13$~TeV are discussed. Additionally, new studies of correlations between $\Lambda_{\rm c}^+$ and protons in pp collisions at $\sqrt{s} = 13.6$~TeV, utilizing the new data samples collected from LHC Run~3, are presented.

}

\maketitle

\section{Introduction and physics motivation}
\label{intro}

Heavy quarks are produced in initial hard scatterings in nucleus–nucleus collisions predominantly due to their large masses, 
making them effective probes of the Quark--Gluon Plasma (QGP) \cite{ALICE:2022wpn}. 
After hadronization, interactions between charm hadrons and light hadrons in the hadronic phase 
become crucial for understanding final-state effects. Theoretical predictions based on lattice QCD (LQCD) and effective models predict very small D--light meson scattering parameters ($\sim$0.1--0.5 fm), 
much smaller than light--light ($\sim$7--8 fm) or light--strange ($\sim$1--2 fm) systems. 
Some channels, such as D--$\pi$ with isospin 1/2, are missing in LQCD data, 
yet are closely connected to the structure of the $\mathrm D_{\mathrm s0}^{*}(2300)$ \cite{Guo:2018kno}\cite{Mohler:2012na}\cite{Liu:2012zya}. 
Thus, experimental constraints are urgently needed.  
Moving on to the charm baryon sector, an important objective is the search for possible charmed hypernuclei.  
The lightest candidate, often referred to as the ``c-deuteron,'' would be a bound state of a 
$\Lambda_{\rm c}^{+}$ baryon and a neutron, analogous to the deuteron and hypertriton.  
Theoretical studies suggest that such a state could exist only if the $\Lambda_{\rm c}^{+}$--nucleon 
interaction is sufficiently attractive. 
Recent LQCD studies suggest attraction in the both $^{1}S_{0}$ and $^{3}S_{1}$ states\cite{Dover:1977jw}\cite{Haidenbauer:2017dua}. Femtoscopy provides direct access to such interactions through two-particle correlations. 
This article reports ALICE results on D--hadron femtoscopy in Run~2 and the first $\Lambda_{\rm c}^{+}$--proton correlation measurement in Run~3.

\section{Analysis strategy}

The femtoscopy technique is employed to study the strong interaction between charm hadrons and light hadrons 
via the two-particle correlation function $C(k^{*})$. 
Experimentally, the correlation function is constructed as the ratio between the same-event and mixed-event 
pair distributions
\begin{equation}
  C(k^{*}) = \mathcal{N} \, \frac{N_{\text{same}}(k^{*})}{N_{\text{mixed}}(k^{*})} ,
\end{equation}
where $N_{\text{same}}(k^{*})$ and $N_{\text{mixed}}(k^{*})$ denote the distributions of particle pairs from the same 
and mixed events, respectively, and $\mathcal{N}$ is a normalization factor . Here $k^{*}$ is the relative momentum in the pair rest frame \cite{Koonin:1977fh}\cite{Pratt:1990zq}. Theoretically, the correlation function can be expressed as an integral over the source emission function 
$S(\vec{r}^{*})$ and the squared relative wave function $\psi(\vec{r}^{*}, \vec{k}^{*})$ of the particle pair
\begin{equation}
  C(k^{*}) = \int d^{3}r^{*} \, S(\vec{r}^{*}) \, \left| \psi(\vec{r}^{*}, \vec{k}^{*}) \right|^{2} .
\end{equation}
The shape of $C(k^{*})$ reflects the character of the interaction: values above unity indicate attraction, below unity repulsion. Combined with fitted scattering parameters or comparison to theoretical potential models, it can also provide information on a possible bound state. The dependence of $C(k^{*})$ on the source size is constrained in a data-driven way using proton–proton correlations~\cite{Fabbietti:2020bfg}.

For the analysis of $\Lambda_{\rm c}^{+}$--p femtoscopy was studied using approximately 
$5.8~\mathrm{pb}^{-1}$ of minimum-bias pp collisions at $\sqrt{s}=13.6$~TeV collected with ALICE. 
$\Lambda_{\rm c}^{+}$ candidates were reconstructed via the $\Lambda_{\rm c}^{+} \rightarrow \rm pK^{-}\pi^{+}$ decay. A multi-class BDT was used for efficient selection and to separate prompt from non-prompt contributions. After proton selection, same-charge $\Lambda_{\rm c}^{+}$–p pairs were extracted in $k^{*}$ intervals from same- and mixed-event samples.

\section{Results from ALICE Run 2}

\begin{figure}[h]

\centering
\includegraphics[width=5.5 cm,clip]{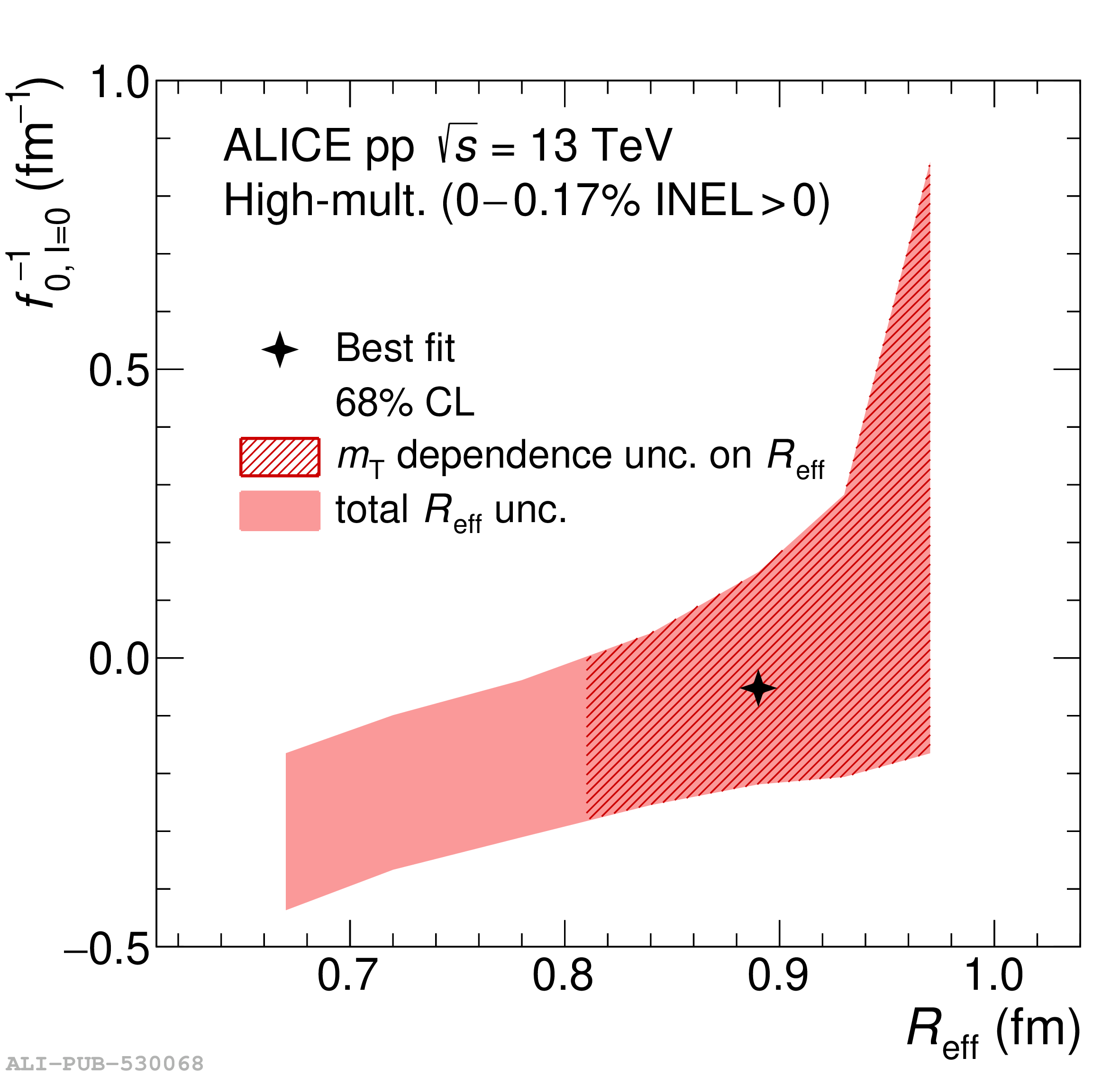}
\includegraphics[width=5.15 cm,clip]{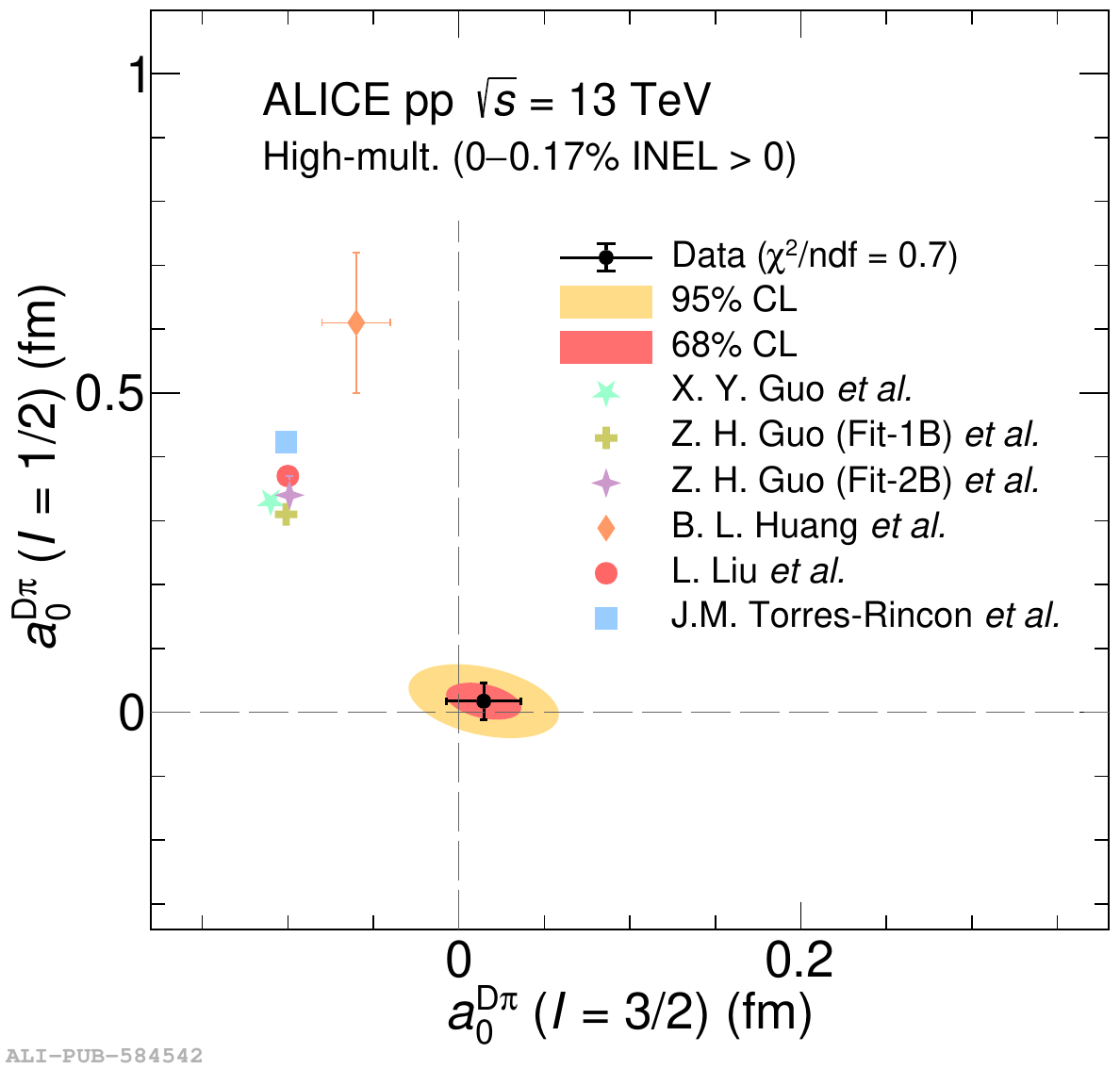}

    \caption{Left: Constraints on the scattering length in the $\rm I=0$ channel for the $\rm D^{+}$--p correlation function 
    in pp collisions at $\sqrt{s}=13$~TeV. The band shows the uncertainty due to the source size determination \cite{ALICE:2022enj}. 
    Right: Extracted scattering parameters for the $\rm D$--$\pi$ system compared to various theoretical predictions. 
    The results indicate vanishing scattering lengths within uncertainties, with a significant tension in the 
    $\rm I=1/2$ channel \cite{ALICE:2024bhk}.}
    \label{fig:Dp_Dpi}
\end{figure}

In ALICE Run~2, femtoscopy correlations were measured between charm mesons and light hadrons \cite{ALICE:2022enj}.  
In the left panel of Fig. \ref{fig:Dp_Dpi}, the $\rm D^{+}$--p correlation function provides the first experimental constraint on the scattering length in the $\rm I=0$ channel, 
suggesting either a shallow attractive interaction without a bound state, 
or a strong attraction in the case of bound-state formation \cite{ALICE:2022enj}.  The $\rm D$--$\pi$ and $\rm D^{*}$--$\pi$ systems were also studied \cite{ALICE:2024bhk}. 
In both cases, the extracted scattering lengths are close to zero, 
consistent with vanishing interaction parameters. 
However, a significant tension with theoretical models is observed, in particular in the $\rm I=1/2$ channel of the $\rm D$--$\pi$ system, as shown in the right panel of Fig. \ref{fig:Dp_Dpi}.  
The $\rm D^{*}$--$\pi$ results are in agreement with those of $\rm D$--$\rm \pi$, as expected from heavy-quark spin symmetry \cite{ALICE:2024bhk}. These measurements represent the first femtoscopic constraints on charm--light interactions, 
and they pave the way for more precise studies with the upgraded ALICE detector in Run~3.  

\section{Results from ALICE Run 3}

The measured $\Lambda_{\rm c}^{+}$--p pairs can be decomposed into several components, and consequently the correlation 
function is expressed as a sum of contributions from different sources 
\begin{equation}
  \begin{split}
    C(k^{*}) = \mathcal{N} \times C_{\text{back}}(k^{*}) \times \Big[ &
    \lambda_{\Lambda_{\rm c} \rm p}^{\text{non-prompt}} C_{\Lambda_{\rm c} \rm p}^{\text{non-prompt}}(k^{*}) \\
    &+ \sum_{i} \lambda_{\Sigma_{\rm c} \rm p}^{\text{prompt, not direct}} C_{\Sigma_{\rm c} \rm p}(k^{*})
     + \lambda_{\Lambda_{\rm c} \rm p}^{\text{prompt, direct}} C_{\Lambda_{\rm c} \rm p}^{\text{direct}}(k^{*}) 
     + \lambda_{\text{flat}} \Big] .
  \end{split}
\end{equation}

\begin{figure}[h]

\centering
\includegraphics[width=6.45 cm,clip]{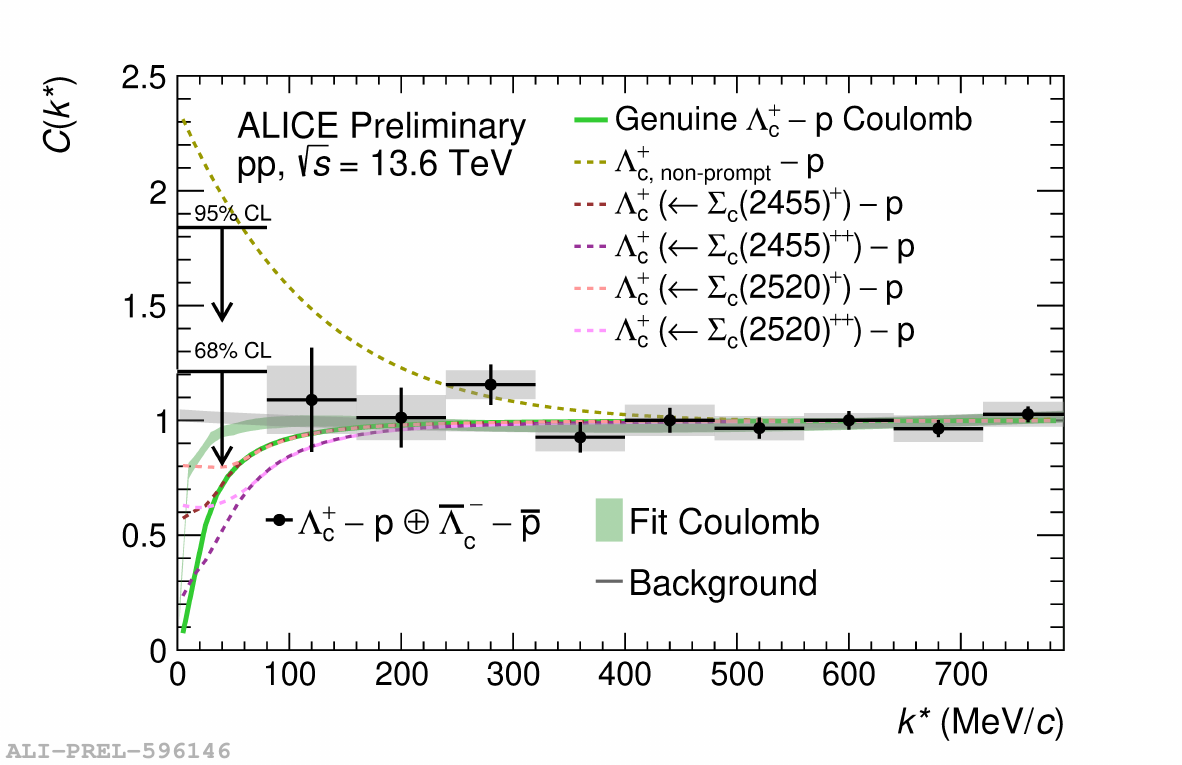}
\includegraphics[width=6.45 cm,clip]{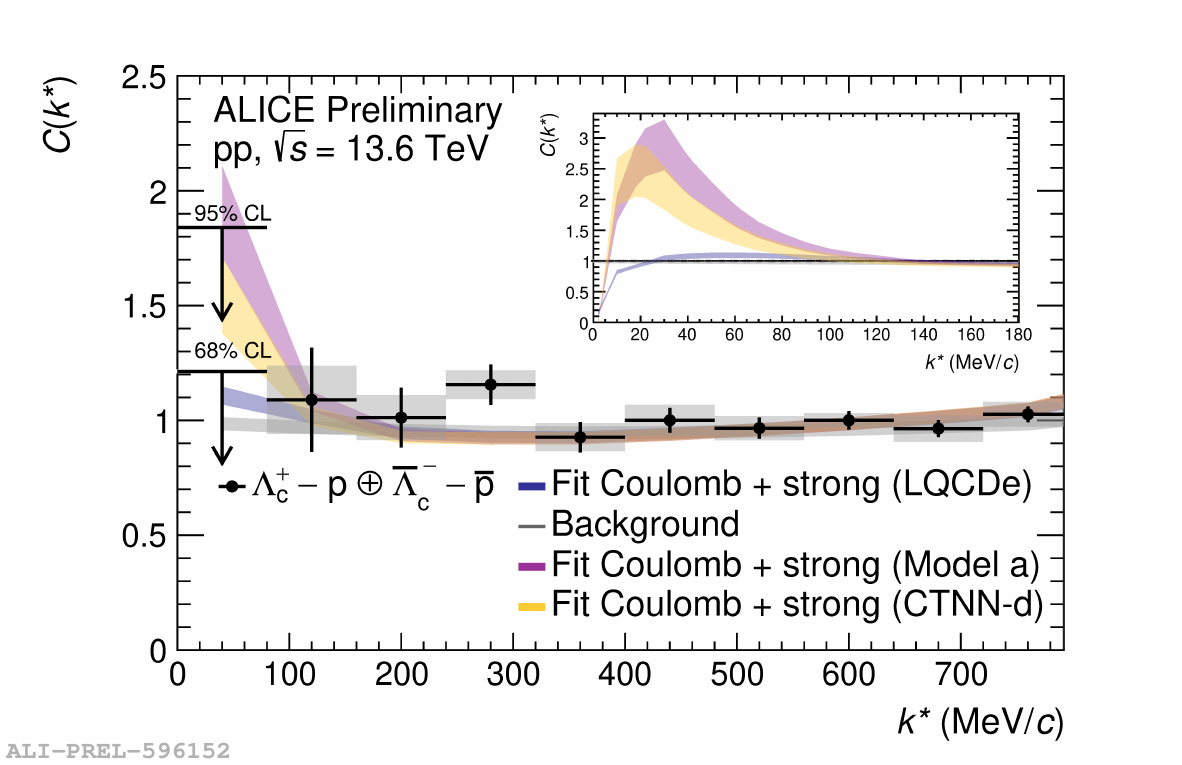}

    \caption{Correlation function of $\Lambda_{\rm c}^{+}$--p pairs in pp collisions at $\sqrt{s}=13.6$~TeV: 
    left, decomposition into different sources; right, comparison with theoretical models of the strong interaction.}
    \label{Lcp_decomposition}  
\end{figure}

The first contribution originates from non-prompt $\Lambda_{\rm c}^{+}$ baryons, produced in weak decays of b-hadrons. 
Since the average decay length of B-hadrons ($\sim$400--500~$\mu$m) is huge compared to the source size, 
the effect of this component on the correlation function is minor. 
The non-prompt fraction is estimated using a $\rm \chi^{2}$ minimization approach \cite{ALICE:2021mgk}, and its correlation function is 
modeled with \textsc{Pythia}~8 simulations in Color Reconnection (CR) Mode~2 \cite{Skands:2014pea}.  The second contribution arises from the feed-down of excited $\Sigma_{\rm c}$ states, in particular the 
$\Sigma_{c}(2455)$ and $\Sigma_{\rm c}(2520)$. 
The feed-down fraction is constrained using published ALICE data \cite{ALICE:2021rzj}. 
Due to the lack of theoretical scattering calculations for these states, only the Coulomb interaction is 
considered in the present analysis.  Finally, the prompt and direct $\Lambda_{\rm c}^{+}$--p pairs represent the genuine strong-interaction component, 
accounting for approximately 20\% of the total. 
The decomposition of the correlation function into these contributions is shown in  the left panel of Fig.~\ref{Lcp_decomposition}, where all components are consistently modeled within the analysis.  

The measured $\Lambda_{\rm c}^{+}$--p correlation function is compared to theoretical models that include 
both Coulomb and strong interactions. 
These models incorporate the proper contributions from experimental data and differ in their assumptions 
about the strength of the strong potential \cite{Haidenbauer:2017dua}\cite{Haidenbauer:2020kwo}\cite{Vidana:2019amb} \cite{Maeda:2015hxa}:  

\begin{itemize}
  \item \textbf{LQCD-e:} Coulomb interaction plus a shallow attractive strong potential.  
  \item \textbf{Model a:} Coulomb interaction plus a large attractive strong potential.  
  \item \textbf{CTNN-d:} Coulomb interaction plus a large attractive strong potential, 
        including the possibility of a deuteron-like bound state.  
\end{itemize}

The comparison between data and models is shown in the right panel of Fig.~\ref{Lcp_decomposition}. 
The measured correlation function is evaluated with the same bin width as the models, 
while the inset panel displays the model predictions with finer binning.  
The results indicate that the data favor a shallow attractive strong interaction at the 
$1\sigma$ confidence level, If confirmed with larger data samples, this would disfavor the formation of charmed hypernuclei.

\section{Summary and outlook}

In summary, a first set of measurements of charm hadron interactions have been performed with ALICE. 
In Run~2, femtoscopic studies focused on correlations between D mesons and light-flavor hadrons, 
providing the first experimental constraints in the charm sector. 
With the upgraded Inner Tracking System (ITS~2) and the increased integrated luminosity in LHC Run~3, significantly higher statistics are now available to explore charm baryon–proton interactions. 
The first preliminary $\Lambda_{\rm c}^{+}$--p results indicate a shallow attractive strong interaction 
at the $1\sigma$ confidence level, with further improvements expected from the triggered 
data samples collected in 2023 and 2024.  


Looking beyond Run 3 and 4, ALICE~3 will offer unique opportunities to reconstruct charmed hypernuclei, 
opening new frontiers in hypernuclear physics and enabling precision studies of the strong interaction 
in systems with open charm.  
\bibliography{template} 
%

\end{document}